\documentstyle[iopconf1]{article}

\def\apj#1{{\em Astrophys. J.} {\bf #1}}
\def\mn#1{{\em Mon. Not. R. astr. Soc.} {\bf #1}}
\def\aa#1{{\em Astron. Astrophys.} {\bf #1}}
\def\nat#1{{\em Nature} {\bf #1}}
\def\apjs#1{{\em Astrophys. J. Suppl.} {\bf #1}}
\def\etal{{\it et al\/}\ }
\def\Mpc{$h^{-1}$~{\rm  Mpc}}
\def\hmpc{$h$~{\rm  Mpc$^{-1}$}}

\begin{document}

\title{REGULARITY OF THE LARGE-SCALE STRUCTURE OF THE UNIVERSE}

\author{Jaan Einasto  \footnote{E-mail:einasto@aai.ee}}

\affil{ Tartu Observatory, EE-2444 T\~oravere, Estonia}

\beginabstract
Recent analysis of the distribution of clusters of galaxies is
reviewed. Clusters of galaxies located in rich superclusters form a
quasiregular lattice. The power spectrum of clusters of galaxies has a
sharp peak at wavelength $\lambda_0=120$~\Mpc\ corresponding to the
lattice step.  The peak in the spectrum may be due to processes during the
inflationary stage of the structure evolution.  
\endabstract

\section{Introduction}

Galaxies and systems of galaxies are formed due to initial density
perturbations of different scale. Short perturbations of a scale of
several Megaparsecs give rise to the formation of galaxies, medium
scale perturbations lead to the formation of clusters of galaxies.
Perturbations of a characteristic scale of $\sim 100$\ \Mpc\ are
related to superclusters of galaxies. Still larger perturbations have
a lower amplitude and modulate densities and masses of smaller systems
(Frisch \etal 1995).

We use the term superclusters of galaxies for the largest relatively
isolated density enhancements in the Universe (Einasto \etal
1997b). Superclusters consist of filamentary chains of galaxies,
groups and clusters of galaxies, and voids between them.
Superclusters are not completely isolated in space. Galaxy and cluster
filaments connect neighbouring superclusters to a single network.
Superclusters and voids form a continuous web which extends over the
whole observable part of the Universe.

Let us accept the inflationary paradigm, and assume that the Universe
is dominated by cold dark matter (CDM) and that initial density
perturbations have a Gaussian distribution. Under these assumptions
large-scale high- and low-density regions should be randomly
distributed.  It was a great surprise when Broadhurst \etal (1990)
found that the distribution of high-density regions in a small area
around the northern and southern Galactic pole is fairly regular:
high- and low-density regions alternate with a rather constant step of
128~\Mpc.  The power spectrum of galaxies of the Broadhurst survey has
a sharp peak on the scale equal to the step of supercluster network along
the line of sight.  Recent analysis of deep galaxy and cluster samples
has shown that the power spectrum of these objects has also a spike on
similar scale.  The presence of such a spike is difficult to explain.

In this review I shall give a summary of recent work on the distribution
of galaxies and clusters of galaxies on large scales. The distribution
is characterised quantitatively by the power spectrum of density
perturbations of these objects. Our goal is to explain the spike in
the observed power spectrum.

\section{The distribution of high-density regions}

Here we shall discuss the distribution of high-density regions in the
Universe.  High-density regions of the Universe can be located by
clusters in very rich superclusters.

The Abell-ACO cluster sample (Abell 1958, Abell, Corwin and Olowin
1989) is presently the largest and deepest sample of extragalactic
objects which covers the whole sky out of the Milky Way zone of
avoidance. Thus the study of the distribution of Abell-ACO clusters is
of special interest. We have used the compilation of all available
redshifts of Abell-ACO clusters by Andernach and Tago (1998, in
preparation), and the supercluster catalogue by Einasto \etal (1997b)
based on this compilation.

The distribution of Abell-ACO clusters shows that very rich
superclusters form a fairly regular lattice (Einasto \etal 1997b,
 1998d).  In Figures~1 and 2 we give the distribution of superclusters
consisting of 8 and more clusters of galaxies. A regular lattice is
very well seen in the southern Galactic hemisphere.  In the northern
hemisphere the regularity can be best seen if we use a thin slice as
in Figure~1.

\begin{figure*}[t]
\vspace*{6cm}
\caption{ Distribution of clusters in high-density regions in
supergalactic coordinates. A 100~\Mpc\ thick slice is shown to avoid
projection effects. Only clusters in superclusters with at least 8 members
are plotted.  The supergalactic $Y=0$ plane coincides almost exactly with
the Galactic equatorial plane; Galactic zone of avoidance is marked by
dashed lines.  }
\includegraphics{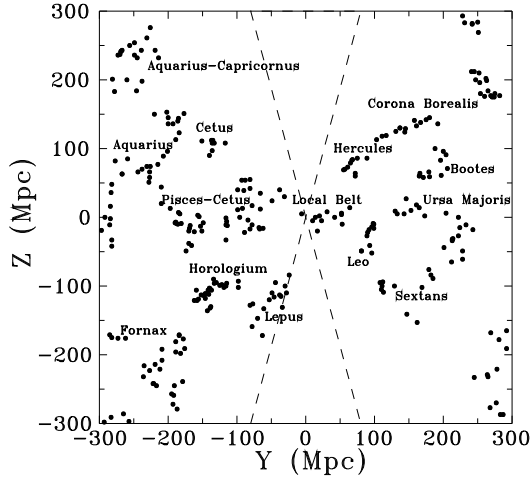}
\label{fig1}
\end{figure*}

Clusters of galaxies in less rich superclusters and isolated clusters
are basically located in walls between voids formed by very rich
superclusters, see Figure~2. Their distribution complements the
distribution of very rich superclusters, together they form a fairly
regular lattice. Cells of this lattice have a characteristic diameter
about 120~\Mpc.

\begin{figure*}[t]
\vspace*{6cm}
\caption{ Distribution of Abell-ACO and APM clusters in high-density
regions in supergalactic coordinates. Abell-ACO clusters in superclusters
with at least 8 members are plotted; APM clusters are plotted if located in
superclusters of richness 4 and higher. The supergalactic $Y=0$ plane
coincides almost exactly with the Galactic equatorial plane and marks the
Galactic zone of avoidance. In the left panel some superclusters overlap
due to projection, in the right panel only clusters in the southern
Galactic hemisphere are plotted and the projection effect is small.  }
\includegraphics{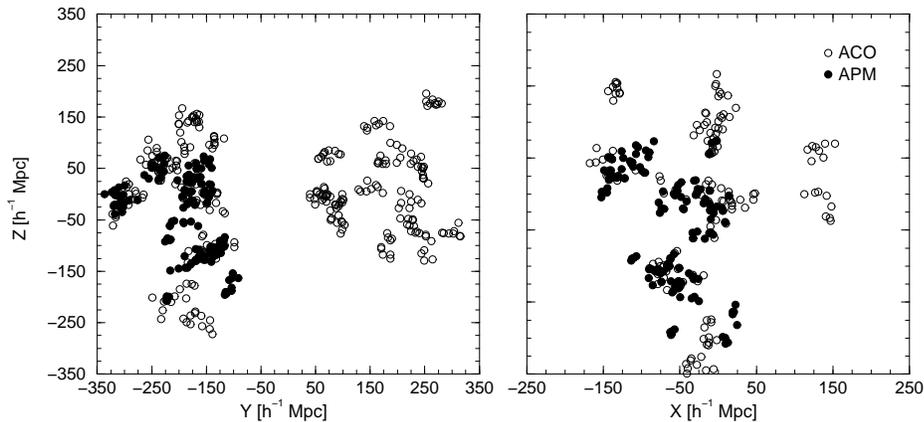}
\label{fig2}
\end{figure*}

A regular distribution of objects in high-density regions has been
detected so far only for Abell-ACO clusters of galaxies. It has been
argued that this sample is distorted by selection effects and that the
regular location of clusters may be spurious. To check this possibility we
compare the distribution of Abell-ACO clusters with that of APM clusters. 
Results for clusters in rich superclusters are shown in Figure~2. This
comparison shows that all rich superclusters are well seen in both cluster
samples. There is a systematic difference only in the distribution of more
isolated clusters since the APM sample contains some clusters also in
less-dense environment (in voids defined by Abell-ACO clusters) as seen
in Figure~2. 

Figure~2 demonstrates also why the overall regularity of the distribution
of APM clusters is much less pronounced. The APM sample covers only a
small fraction of the volume of the Abell-ACO cluster sample: APM
sample is located in the southern hemisphere and even here it covers a 
smaller volume. Thus it is not surprising that the grand-design of the
distribution of high-density regions is not well seen in the APM sample.

This comparison of the distribution of Abell-ACO and APM clusters
shows that Abell-ACO clusters are good tracers of the large-scale
distribution of high-density regions. Presently the Abell-ACO sample
provides the best candidate to a fair sample of the Universe.

\section{The power spectrum of galaxies and clusters of galaxies}

Einasto \etal (1998a) have analysed recent determinations of power
spectra of galaxies and clusters of galaxies; a summary is given in
Figure~1.  Here power spectra of different objects are vertically
shifted to coincide amplitudes near the wavenumber $k=2\pi /\lambda =
0.1$~\hmpc.  On medium scales spectra have a negative spectral index;
near the scale $l\approx 120$ \Mpc\ or wavenumber $k=2\pi/l = \approx
0.05$~\hmpc\ spectra have a turnover; on very large scales the
spectral index approaches the theoretically predicted value $n=1$.

\begin{figure*}[t]
\vspace*{5cm}
\caption{Power spectra of galaxies and clusters of galaxies scaled to
match the amplitude of the 2-D APM galaxy power spectrum (Einasto
\etal 1998a). Spectra are shown as smooth curves.  Bold lines show
spectra for clusters data: short dashed line for Abell-ACO clusters
according to Einasto \etal (1997a), long-dashed line according to
Retzlaff \etal (1998), dot-dashed line for APM clusters (Tadros \etal
1998); thin lines for galaxies: short dashed line for IRAS galaxies
(Peacock 1997), long-dashed line for SSRS-CfA2 galaxy survey (da Costa
\etal 1994), dot-dashed line for LCRS (Landy \etal 1996). The solid
thick line shows the mean power spectrum, the dashed thick line
indicates the power spectrum derived from the APM 2-D galaxy
distribution (Peacock 1997, Gazta\~naga and Baugh 1998).  }  
\includegraphics{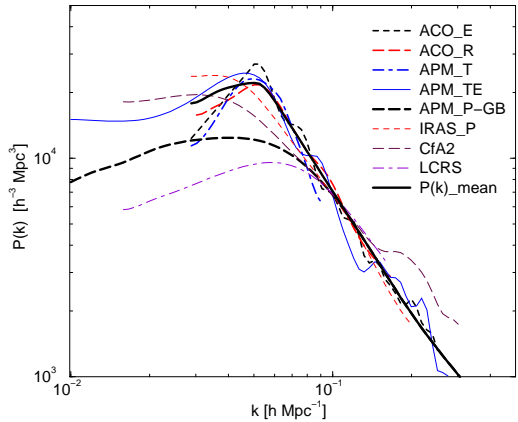}
\label{fig3}
\end{figure*}

A closer inspection shows that after scaling samples have different
amplitudes near the turnover. Cluster samples as well as APM and
SSRS-CfA2 galaxy samples have a high amplitude near the maximum.
Spectra of cluster samples have here a sharp spike. In contrast, IRAS
and LCRS galaxy samples have a much shallower transition of lower
amplitude; the power spectrum calculated from the 2-D distribution of
APM galaxies has also a shallow maximum.  The reason for this
difference is not fully clear. It is possible that the spatial
distribution of different objects is different. On the other hand, it
is not excluded that differences may be partly caused by disturbing
effects in data reduction. For instance, the window function of the
LCRS is very broad in Fourier space which can smear out the spike in
the power spectrum.

We have formed two mean power spectra of galaxies. One mean spectrum
is based on cluster samples and APM and SSRS+CfA2 galaxy samples, it
has a relatively sharp maximum at $k=0.05$~\hmpc, and a power law
behaviour on medium scales with index $n=-1.9$. Cluster samples cover
a large volume where rich superclusters are present; we call this
power spectrum as characteristic for populations in high-density
regions, and designate the spectrum $P_{HD}(k)$.  The second mean
power spectrum was derived from spectra of the LCRS sample, the IRAS
galaxy sample, and the APM 2-D sample. These catalogs sample regions
of the Universe characteristic for medium-rich superclusters, we
designate this spectrum $P_{MD}(k)$. Both mean power spectra are shown
on Figure~3.

\section{Reduction of mean power spectra to matter}

It is well known that the evolution of matter in low- and high-density
regions is different.  Gravity attracts matter toward high-density
regions, thus particles flow away from low-density regions, and
density in high-density regions increases until contracting objects
collapse and form galaxies and clusters of galaxies.  The collapse
occurs along caustics (Zeldovich 1970).  Bond, Kofman and Pogosyan
(1996) demonstrated that the gravitational evolution leads to the
formation of a web of high-density filaments and under-dense regions
outside of the web.  The contraction occurs if the over-density
exceeds a factor of 1.68 in a sphere of radius $r$ which determines
the mass of the system or the galaxy (Press \& Schechter 1974).  In a
low-density environment the matter remains primordial.  This
difference between the distribution of the matter and galaxies is
called biasing. The gravitational character of the density evolution
of the Universe leads us to the conclusion that galaxy formation is a
threshold phenomenon: galaxies form in high-density environment, and
in low-density environment the matter remains in primordial dark form.

The power spectrum of clusters of galaxies has an amplitude larger
than that for galaxies and matter.  We can define the bias
parameter $b_c$ through the power spectra of mass, $P_m(k)$, and of
the clustered matter, $P_c(k)$,
$$
P_c(k) = b_c^2(k) P_m(k),
\eqno(1)
$$
where $k$ is the wavenumber in units of $h$~Mpc$^{-1}$, and the Hubble
constant is expressed as usual, $100~h$ km~s$^{-1}$~Mpc$^{-1}$. In
general the biasing parameter is a  function of wavenumber $k$.

Gramann \& Einasto (1992), Einasto \etal (1994) and Einasto \etal
(1998b) investigated the power spectra of matter and simulated
galaxies and clusters. They showed that the exclusion of non-clustered
matter in voids from the population shifts the power spectrum of the
clustered matter in high-density regions to higher amplitude. The bias
factor calculated from equation (1) is surprisingly constant, and can
be found from the fraction of matter in the high-density clustered
population, $F_c= N/N_{tot}$,
$$ 
b_c = 1/F_c.
\eqno(2)
$$

\begin{figure}[ht]
\vspace*{5cm}
\caption{ Left: Power spectra of simulated galaxies. The solid bold
line shows the spectrum derived for all test particles (the matter
power spectrum); dashed and dotted bold lines give the power spectrum
of all clustered particles (sample  Gal-1), and clustered galaxies in
high-density regions (sample Clust). Thin solid and dashed lines show
the power spectra of samples of particles with various threshold
densities and sampling rules, see Table~1 and text for details. Right:
the biasing parameter as a function of wavenumber, calculated from
definition eqn. (1). Samples and designations as in left panel.
}
\includegraphics{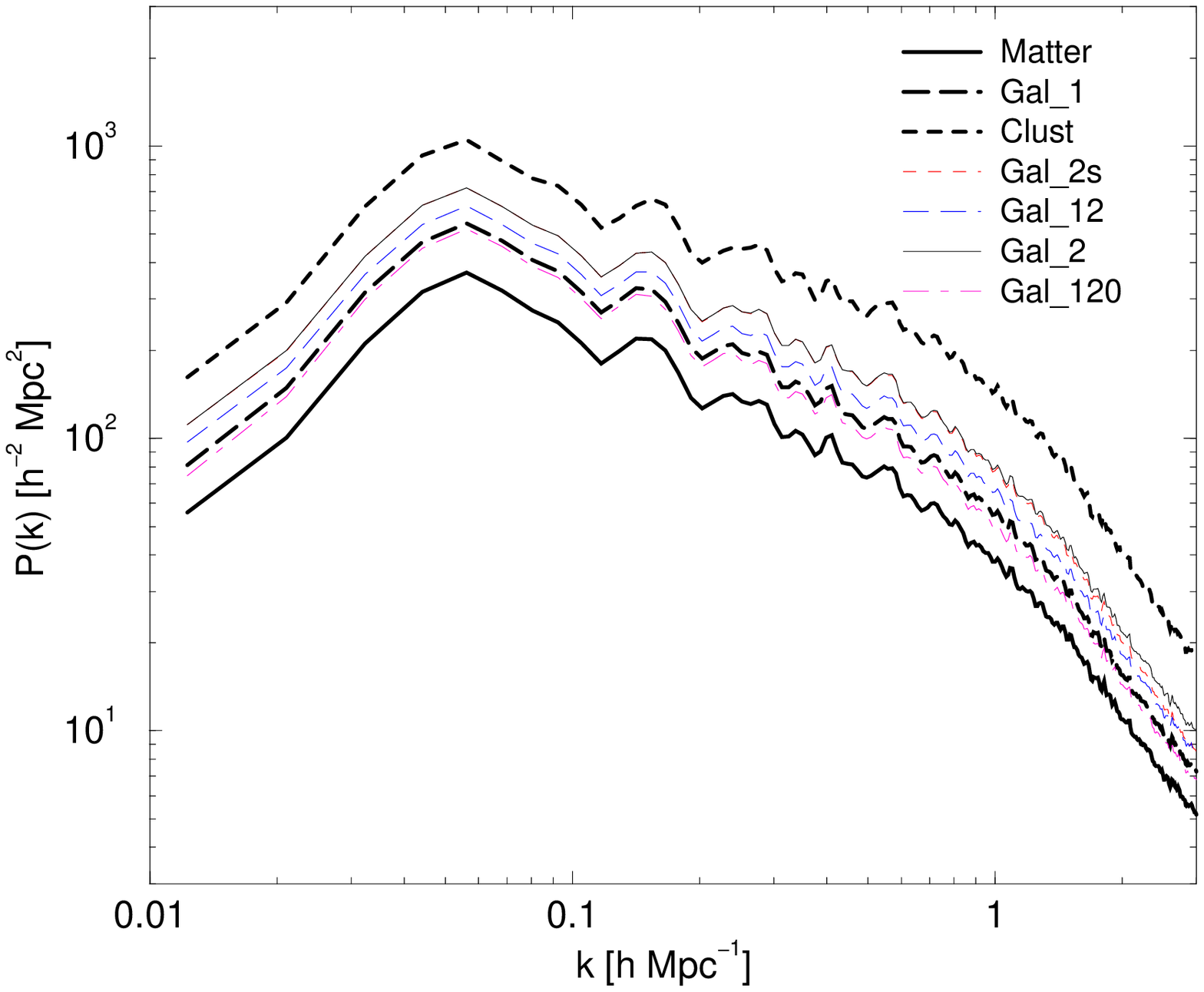} 
\includegraphics{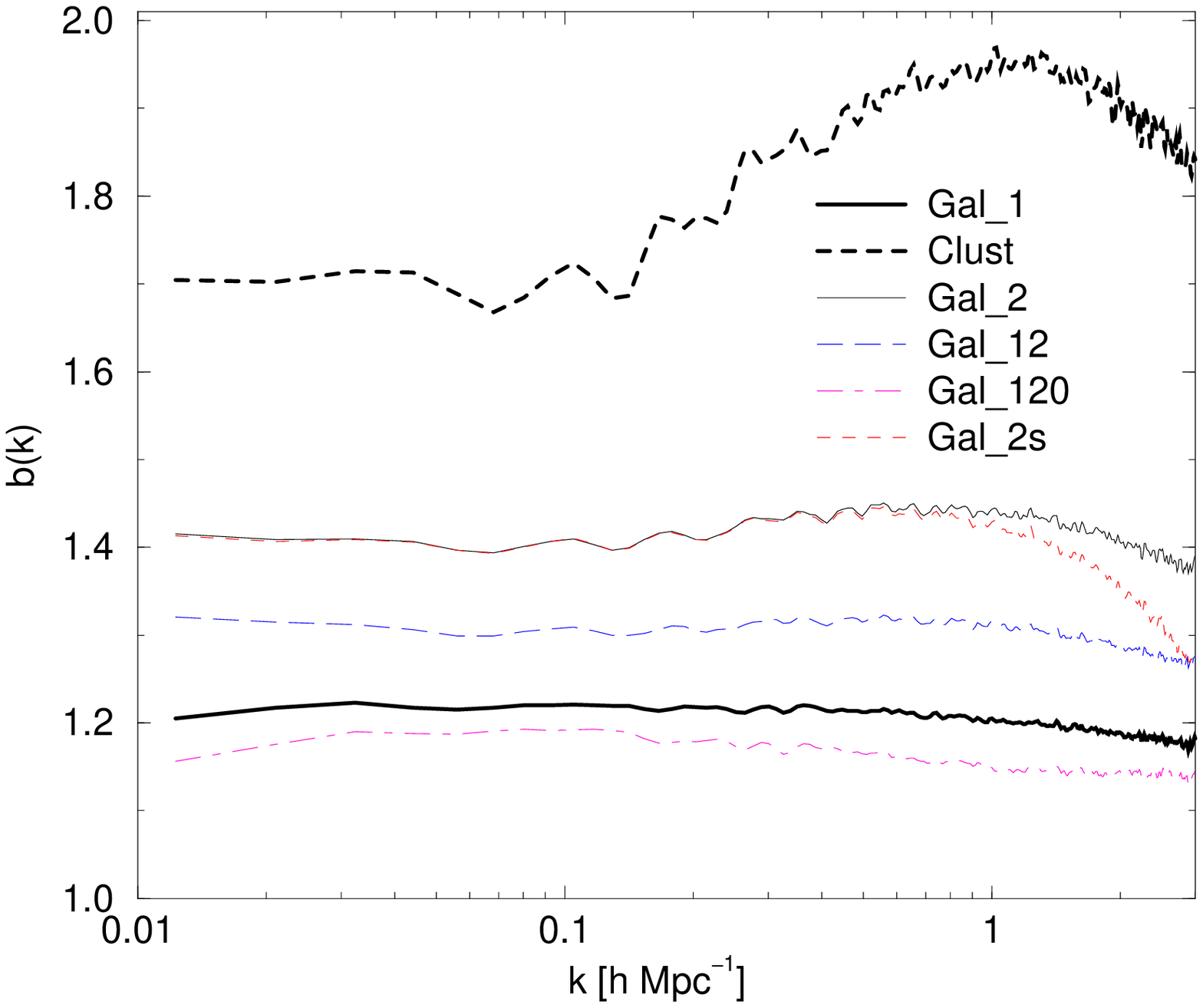}
\label{figure4}
\end{figure} 

In Figure~4 we show the power spectra and related biasing factors for
several simulated galaxy populations, calculated using various
threshold densities to separate the unclustered matter in voids and
the clustered matter associated with galaxies in high-density regions
(Einasto \etal 1998b).  In calculating densities a small smoothing
length $\approx 1$~\Mpc\ was used; in this case the clustered matter
associated with galaxies and their dark haloes is not mixed with
non-clustered matter in voids.  This analysis shows that biasing
factor is almost independent of the scale, and can be calculated from
equation (1). The flow of matter from low-density regions was studied
by Einasto \etal (1994) and Einasto \etal (1998b). They showed that
this flow depends slightly on the density parameter of the structure
evolution model. In the standard CDM model the speed of void
evacuation is slightly faster than in open CDM models and spatially
flat CDM models with cosmological constant. The present epoch of the
model was found using the calibration through the $\sigma_8$ parameter
-- rms density fluctuations within s sphere of radius 8~\Mpc. This
parameter was determined for the present epoch for galaxies,
$(\sigma_8)_{gal} = 0.89 \pm 0.05$. Using this estimate and CDM models
with lambda term and density parameter $\Omega_0 \approx 0.4$ Einasto
\etal (1998b) get $(\sigma_8)_m = 0.68 \pm 0.06$, $F_{gal} = 0.75 \pm
0.05$ and $b_{gal} = 1.3 \pm 0.1$.

\section{Primordial power spectrum of matter}

The semi-empirical power spectrum of matter, found from galaxy and
cluster surveys and reduced to the matter, can be used to compare with
theoretical models.  Such analysis has been done by Einasto \etal
(1998c). CDM models with and without cosmological constant, mixed dark
matter models, and open CDM models were used for comparison. Models
were normalised to four year COBE observations; a Hubble parameter $h
= 0.6$ was used; the density in baryons was taken $\Omega_{b} = 0.04$
(in units of the critical density); the density parameter
$\Omega_0=\Omega_b + \Omega_{DM}$ was varied, the model was kept flat
using cosmological constant $\Omega_{\Lambda}$. In mixed DM models the
density of neutrinos was $\Omega_{\nu}=0.1$; in open models the
density parameter was varied between $\Omega_{0} = 0.2$ and
$\Omega_{0} = 1$.

The observed power spectrum is influenced on small scales by
non-linear evolution of the structure, on these scales the spectrum
was reduced to linear case using analytical models, the best agreement
was achieved by a MDM model with density parameter $\Omega_0 = 0.4$.
In Figure~5 we show the semi-empirical linear power spectra of matter,
which represent galaxy populations in high- and medium-density
regions.  These empirical spectra are compared with analytical power
spectra for MDM models calculated for various density parameter values
from 0.25 to 1.0.

\begin{figure}[ht]
\vspace*{5cm}
\caption{The semi-empirical matter power spectra compared with
theoretical and primordial power spectra. Left: present power spectra;
right: primordial spectra; for MDM models. Solid bold line shows the
linear matter power spectrum found for regions including rich
superclusters, $P_{HD}(k)$; dashed bold line shows the linear power
spectrum of matter $P_{MD}(k)$ for medium dense regions in the
Universe.  Model spectra with $\Omega_{0}= 0.9, ~0.8 \dots ~0.25$ are
plotted with thin lines; for clarity the models with $\Omega_0 = 1.0$
and $\Omega_0 = 0.5$ are drawn with dashed lines.  Primordial power
spectra are shown for the peaked matter power spectrum, $P_{HD}(k)$;
they are divided by the scale-free spectrum, $P(k) \sim k$.  The mean
error of semi-empirical power spectra is about 11~\%.
}
\includegraphics{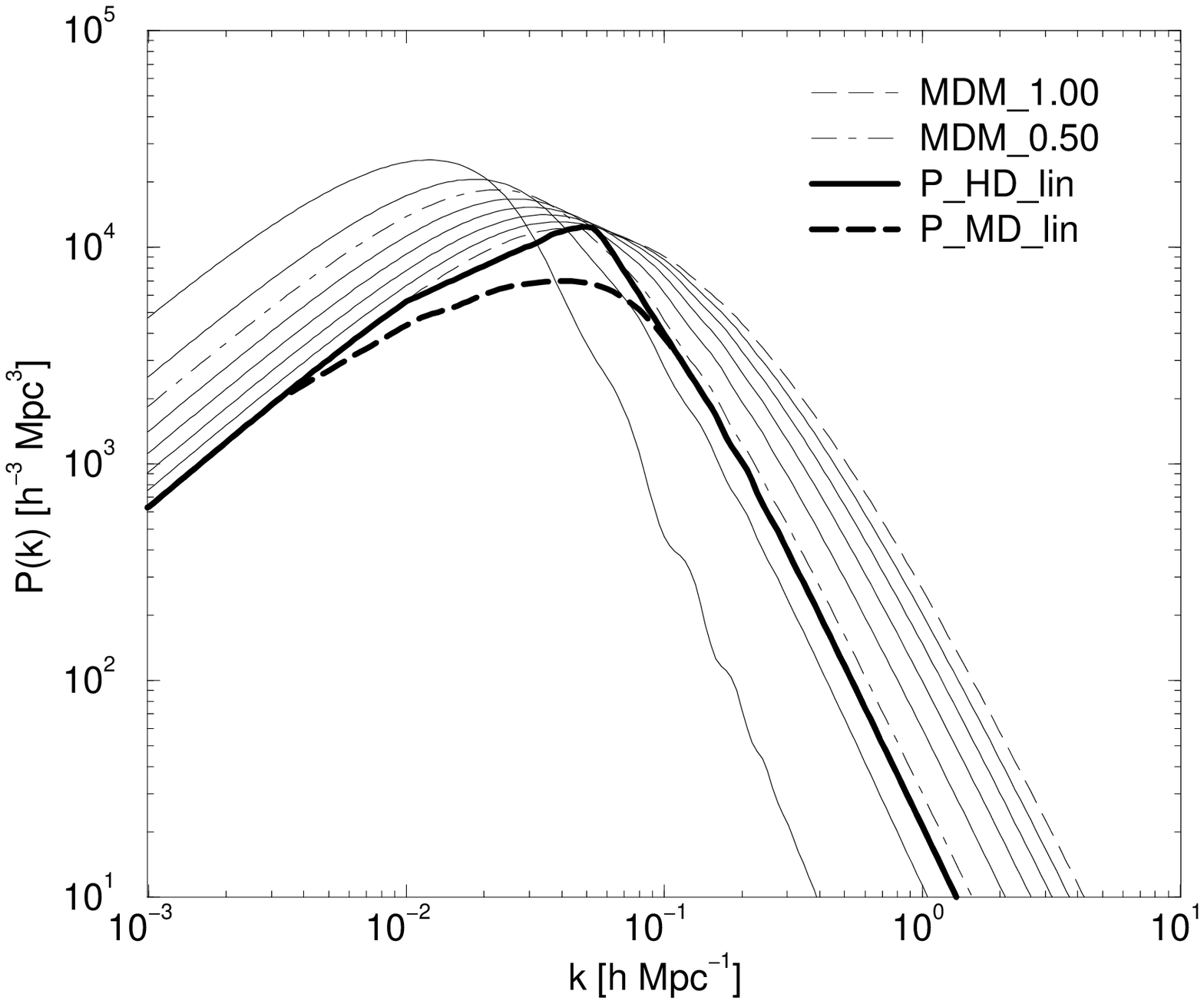} 
\includegraphics{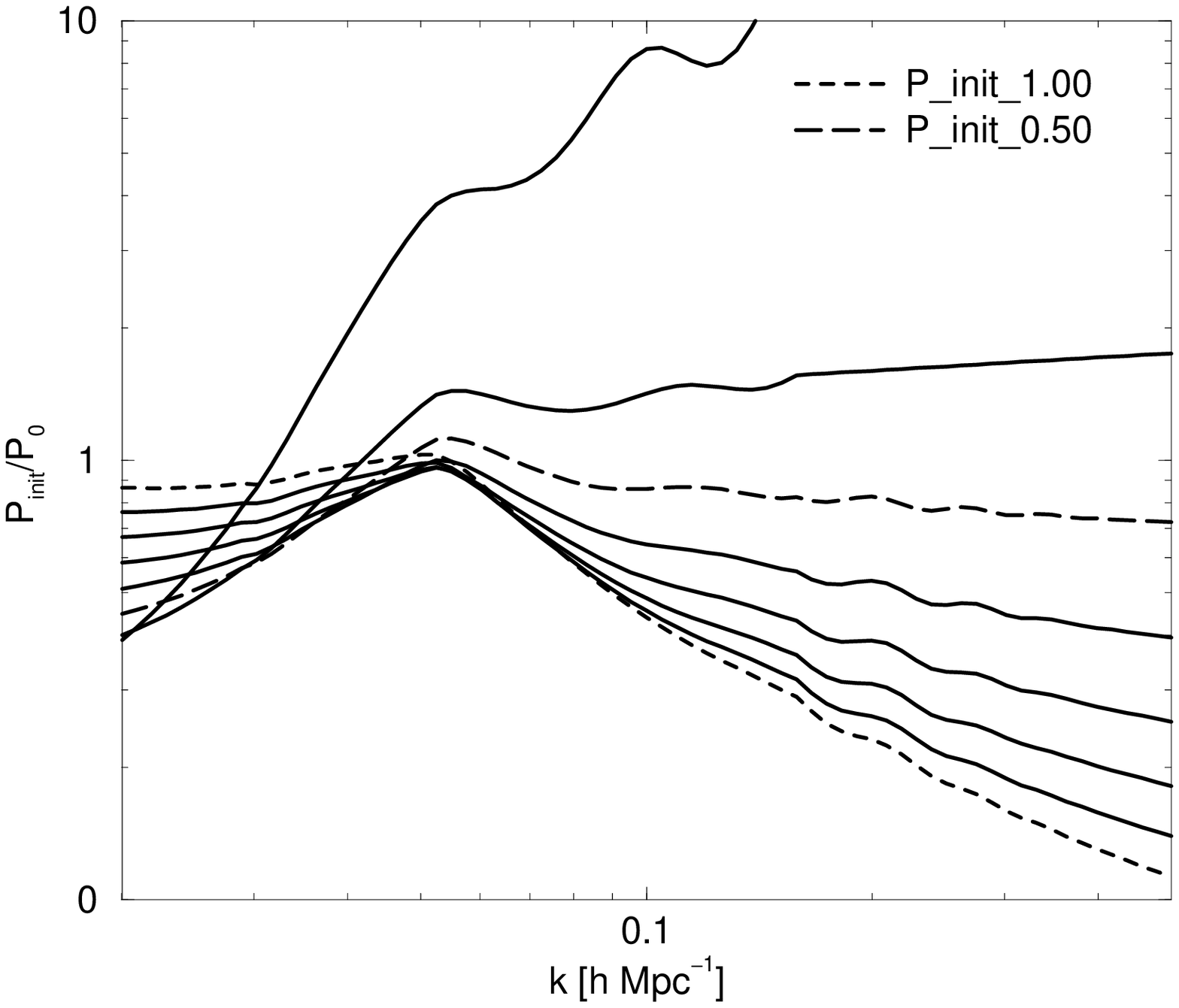}
\label{figure5}
\end{figure}

The comparison of observed and theoretical power spectra gives us also
the possibility to calculate the primordial or initial power spectra
$$ 
P_{init}(k) = P(k)/T^{2}(k),
\eqno(3)
$$ 
where $T(k)$ is the transfer function.  In right panels of Figure~5 we
plot the ratio of the primordial power spectrum to the scale-free
primordial power spectrum, $P_{init}(k)/ P_0(k)$, where $P_0(k) \sim
k$.  Only results for the peaked power spectrum $P_{HD}(k)$ are
shown.  The primordial power spectrum calculated for the shallower
power spectrum $P_{MD}(k)$ is flatter; on large and small scales
it is similar to the spectrum shown in Figure~5. 

The main feature of primordial power spectra is the presence of a
spike at the same wavenumber as that of the maximum of the observed
power spectrum.  On scales shorter than that of the spike the
primordial spectrum can be well approximated by a straight line (on
log--log plot), i.e. by a tilted model with index $n<1$, if $\Omega_0
\geq 0.5$; and $n>1$, if $\Omega_0 <0.5$.  This approximation breaks,
however, down if we consider the whole scale interval. Additionally,
there is a considerable difference in the amplitude of the primordial
spectrum on small and large scales.  For most values of the density
parameter the amplitude on small scales is lower than on large scales;
for very low values of $\Omega_0$ the effect has opposite sign.

In conclusion we can say that it seems impossible to avoid a break in
the primordial power spectrum.  If the peaked power spectrum
$P_{HD}(k)$, based on cluster and deep galaxy data, represents the
spectrum of a fair sample of the Universe, then the break of the
primordial power spectrum is sudden, otherwise it is smooth. The
relative amplitude of the break, and the respective change in the
power index of the spectrum depends on models and cosmological
parameters used. In the framework of the inflationary paradigm the
primordial power spectrum is generated during the inflation. A broken
scale primordial power spectrum suggests that inflation may be more
complicated than thought so far. One possibility for a broken scale
primordial power spectrum has been suggested by Lesgourgues \etal
(1997).  Future deep surveys of galaxies will fix the initial power
spectrum of matter more precisely.  Presently we can say that a
broken-scale primordial power spectrum deserves serious attention.

\section*{Acknowledgments}
I thank H. Andernach, R. Cen, M. Einasto, S.  Gottl\"ober, A.Knebe, 
V. M\"uller, A. Starobinsky, E. Tago and D. Tucker for fruitful
collaboration and permission to use our joint results in this talk.  
This study was supported by the Estonian Science Foundation.

\end{document}